\documentstyle[11pt,epsfig]{article}

\begin{document}
\begin{titlepage}
\begin{center}
February, 1996      \hfill     HUTP-96/A003 \\
\vskip 0.2 in
{\large \bf INTERPOLATION OF NON-ABELIAN LATTICE GAUGE FIELDS}\footnotetext{~}

\vskip .2 in
       {\bf  Pilar Hern\'andez}\\
and \\
       {\bf   Raman Sundrum}
        \vskip 0.3 cm
       {\it Lyman Laboratory of Physics \\
Harvard University \\
Cambridge, MA 02138, USA}
 \vskip 0.7 cm 

\begin{abstract}
We propose a method for interpolating non-abelian lattice gauge fields to the
continuum, or to a finer lattice, which satisfies the properties of
(i) transverse continuity, (ii) (lattice) rotation and translation covariance, 
(iii) gauge covariance, (iv) locality. These are the properties required
for use in our earlier proposal for non-perturbative formulation and
simulation of chiral gauge theories.
\end{abstract}
\end{center}

\end{titlepage}

\section{INTRODUCTION}
 
Continuum interpolations of lattice gauge fields have been considered in
the literature as a way of relating the topological aspects of continuum 
gauge theory to lattice gauge theory, and as an intermediate step in
coupling lattice gauge fields to fermions in a manner that preserves chiral
symmetry \cite{int1} \cite{int2} \cite{kron} \cite{thooft} \cite{us}. 
Recently, we have proposed a method for defining and
simulating chiral gauge theories, with gauge fields on a coarse lattice
coupled to fermions on a finer lattice via an interpolated gauge field
\cite{us}. We showed there, that in order for our method to succeed the 
interpolation procedure must satisfy the properties (i) -- (iv)
described below. The interpolation procedure we
advocated was adapted from 't Hooft's recent suggestion \cite{thooft} 
in the context of vector-like gauge theories.
Unfortunately, this interpolation does not satisfy (i)\footnote{We are
grateful to R. Narayanan for bringing this to our
attention.}. Indeed we know of no
procedure in the earlier literature which satisfies $all$ the four
properties we need. The purpose of this paper is to fill this gap. While
it is possible to cure the problems of the interpolation of ref. \cite{us},
we will give an alternative procedure which is computationally more
efficient. This procedure is very similar to the proposal 
of ref. \cite{kron}, based on the earlier work of L\"uscher \cite{luscher}.
However this earlier proposal  does not satisfy properties (i) and (ii).

Given a four-dimensional
 non-abelian lattice gauge field configuration, $U_{\mu}(s)$, taking
values in gauge group $G$, our procedure will give a continuum gauge field,
$a_{\mu}(x)$. It is an interpolation in the sense that the
parallel transport of $a_{\mu}(x)$ along the links of the lattice will
equal the lattice link variables, $U_{\mu}(s)$. 
The interpolation, $a[U](x)$, has the following properties:

(i) Transverse Continuity: $a_{\mu}$ is differentiable $inside$ each
hypercube of the lattice and its components transverse to the normal of a
boundary  between adjacent hypercubes are continuous across the boundary.

(ii) Rotational and Translational Covariance: If $T$ is a lattice
(improper) rotation
and/or translation of a lattice gauge field, $U$, then there exists a
continuum gauge transformation, $\omega$, such that
\begin{equation}
a[T[U]] = T[a[U]]^\omega.
\end{equation}

(iii) Gauge Covariance: If $\Omega$ is a lattice gauge transformation of
$U$, then there exists a continuum gauge transformation, $\omega$, which
interpolates $\Omega$ (the two agree on lattice vertices), such that
\begin{equation}
a[U^{\Omega}] = a^{\omega}[U].
\end{equation}

(iv) Locality: In ref. \cite{us} we defined a strong form of locality,
namely that $a[U]$ in any hypercube be determined only by $U$ on the
bounding links. Here we shall only demand that the {\it gauge-invariant 
behavior} of $a[U]$ is determined locally. More precisely, we define
locality to mean that (the trace of) any continuum Wilson loop obtained 
from $a[U]$ is
determined only by $U$ on links of hypercubes through which the Wilson loop
passes. We have argued in ref. \cite{us} that our proposal for formulating
lattice chiral gauge theories is only sensitive to the gauge-invariant
behavior of $a[U]$ in the continuum limit. We therefore expect that the
present weaker form of locality is sufficient for $a[U]$ to be applied in
that context.

A spacetime which is a four-dimensional torus is represented by periodic
lattice gauge fields in a flat spacetime.
 The winding number of the interpolated
continuum gauge field on the torus will be the winding number we assign
to the lattice gauge field. This definition is essentially equivalent to 
L\"uscher's \cite{luscher}\footnote{In fact, the two definitions agree
for ``non-exceptional'' gauge configurations as defined by L\"uscher, 
for sufficiently small $\epsilon$.}. 
Our interpolation will not be periodic
in general, since that is impossible for transversely continuous gauge fields 
with non-zero winding number. However for gauge fields with zero net winding
number our interpolation will indeed be periodic. Only such fields are
needed for our lattice formulation of chiral gauge theories, since we
have shown that all physical effects can be obtained from the topologically
trivial sector   using cluster decomposition of the full theory \cite{us}.
For topologically non-trivial gauge fields the interpolation will be
periodic in the $x_{1,2,3}$-directions, but will satisfy 
\begin{equation}
a_{1,2,3}[x_1, x_2, x_3,  L] = a_{1,2,3}^{\omega}[x_1, x_2, x_3, 0],
\end{equation}
for some gauge transformation, $\omega$, defined on the subspace $x_4 = L$,
where $L$ is the length of the torus in the $x_4$-direction. This
represents a well-defined $connection$ on the $torus$.

In section 2 we define our interpolation procedure to the continuum for
the simplest example of non-abelian gauge group, $SU(2)$, proving
properties (i) -- (iv). In section 3, we work out the analog of our
procedure for a compact $U(1)$ gauge field in two dimensions. This is the
simplest example of most of the techniques in section 2. 
Section 4 contains some closing remarks.

\section{The Interpolation Procedure}

For notational simplicity we will confine ourselves to the simplest
non-abelian gauge group, $SU(2)$. Our procedure is straightforwardly
generalized to any non-abelian group without $U(1)$ factors. 
We will work in units of the lattice
spacing, so a lattice point $s$ has integer valued components. 
We can uniquely write each lattice link variable in the form
\begin{equation}
U_{\mu}(s) = e^{i \vec{A}_{\mu}(s).\vec{\tau}}, ~~|\vec{A}_{\mu}(s)| < \pi, 
\end{equation}
where we are neglecting the measure-zero set of lattice fields where at
least one of the link variables equals exactly $-1$. This assignment
defines an obvious logarithm function for group elements different from 
$-1$, 
\begin{equation}
A = -i \log U,  ~~|\vec{A}| < \pi.
\end{equation}

The basic strategy is to 
work separately in each hypercube of the lattice and
choose a complete axial
 gauge fixing scheme \cite{luscher} for the bounding link 
variables of the cell. The resulting link variables, $\overline{U}$, are
interpolated in some smooth way, $\overline{a}[\overline{U}]$,
 to the whole cell, in the same gauge \cite{kron}. A gauge transformation
is then applied to yield the final transversely continuous interpolation.

The lattice gauge field $U$ will first be interpolated into lower dimensional
sublattices, working up to 4-dimensional interpolation. This will help
to ensure transverse continuity across hypercube boundaries. We will prove
(i) -- (iv) in each dimension.

\subsection{1-dimensional interpolation}

On the points along each lattice link, define
\begin{equation}
a_{\mu}(s + t \hat{\mu}) = A_{\mu}(s),  ~~0 \leq t < 1.
\end{equation}
The only nontrivial property to check is gauge covariance, (iii).
Under a 
 lattice gauge transformation, $\Omega$, $U_{\mu}^\Omega(s) =
\Omega(s)^{-1} U_{\mu}(s) \Omega(s + \hat{\mu}) \equiv  e^{i A'}$
we can define a continuum gauge transformation,
\begin{equation}
\omega(s + t \hat{\mu}) = e^{-i t A_{\mu}(s)} \Omega(s) e^{i t
A'_{\mu}(s)},
\end{equation}
which  satisfies
\begin{equation}
a^{\omega}_{\mu}(s + t \hat{\mu}) = A'_{\mu}(s).
\end{equation}

\subsection{2-dimensional interpolation}

Consider a $\mu-\nu$-oriented plane in the lattice,  $\mu < \nu$.
 In each plaquette, $s_{\mu}  
\leq  x_{\mu} \leq s_{\mu} + 1$, $s_{\nu}  
\leq  x_{\nu} \leq s_{\nu} + 1$, we first define, 
\begin{equation}
\overline{a}_{\mu}(x) = - i (x_{\nu} - s_{\nu}) \log U_{\mu\nu}(s), 
~~\overline{a}_{\nu}(x) = 0,
\end{equation}
where $U_{\mu \nu}(s) = U_{\nu}(s)~ U_{\mu}(s + \hat{\nu})~ U^{-1}_{\nu}(s +
\hat{\mu})~ U^{-1}_{\mu}(s)$ is the standard product of link variables
around the plaquette.

This interpolation is not in general transversely continuous across plaquette 
boundaries. We will fix this up by making a gauge transformation, $\omega$, on
$\overline{a}$ so that the result agrees on the bounding links with the
earlier 1-dimensional interpolation made there, $a = A$. As is easily
verified, this demand actually  determines $\omega$
on the bounding links,
\begin{eqnarray}
\omega(s + t \hat{\mu}) &=& e^{i t A_{\mu}(s)}, \nonumber \\
\omega(s + \hat{\mu} + t \hat{\nu}) &=& e^{i A_{\mu}(s)} e^{i t 
A_{\nu}(s + \hat{\mu})}, 
\nonumber \\
\omega(s + t \hat{\nu}) &=& e^{i t A_{\nu}(s)}, \nonumber \\
\omega(s + t \hat{\mu} + \hat{\nu}) &=& e^{- t \log U_{\mu \nu}(s)} 
e^{i A_{\nu}(s)} e^{i t A_{\mu}(s + \hat{\nu})},
\end{eqnarray}
where $0 \leq t \leq 1$. This continuous function from the boundary of the
plaquette to $SU(2)$ is continuously extendable to the whole plaquette
because $SU(2)$ is simply connected (in contrast to gauge groups with compact
U(1) factors, which we discuss in section 3). We
choose an extension of $\omega$ to the whole plaquette as
follows. Define $\omega = e^{\phi}$, where $\phi$ is the solution of the
2-dimensional Laplace equation, with $\phi = \log \omega$ on the plaquette
boundary. This extension to the whole plaquette is only continuous if 
$\omega \neq -1$ on the plaquette boundary, since we
need to stay away from the discontinuity of our log function on $SU(2)$.
This restriction will
hold in all but a measure-zero set of lattice gauge fields, which we can
safely neglect. 

The final step of the interpolation is to define
\begin{equation}
a^{(2)}_{\mu, \nu}(x) = \overline{a}^{\omega}_{\mu, \nu}(x),
\end{equation}
which is transversely continuous because it agrees with the one-dimensional
interpolation previously performed. The superscript $(2)$ reminds us that this
is 2-dimensional interpolation, to save confusion later.
 It is obvious that the interpolation in
each plaquette is determined completely by the values of $U$ on the
bounding links, so (iv) holds. 

Under a general lattice gauge transformation $\Omega$,
$\overline{a}[U^{\Omega}]$ is given by
\begin{eqnarray}
\overline{a}_{\mu} &=&  - i (x_{\nu} - s_{\nu}) \log(\Omega(s)^{-1} 
U_{\mu\nu}(s) \Omega(s)) \nonumber \\
&=& - i \Omega(s)^{-1} \{(x_{\nu} - s_{\nu})  \log U_{\mu\nu}(s) \} \Omega(s), 
\nonumber \\ 
\overline{a}_{\nu} &=& 0,
\end{eqnarray}
which is clearly gauge equivalent to $\overline{a}[U]$. Therefore
$a^{(2)}[U^{\Omega}]$ and $a^{(2)}[U]$ are also gauge equivalent, 
verifying (iii).  

Rotational covariance was only broken by the dependence of $\overline{a}$ on 
the global coordinate frame which determines the  particular complete axial
gauge that $\overline{a}$ satisfies. A different global coordinate frame
obtained by a rotation would produce a different $\overline{a}$, but it is
easy to explicitly construct a gauge transformation that relates it to the
old $\overline{a}$. Thus the old and new $a^{(2)}$ are also gauge
equivalent, so property (ii) holds.

\subsection{3-dimensional interpolation}

Now let us consider a 3-dimensional sublattice of the 4-dimensional lattice.
Unlike 2-dimensions, in 3-dimensions (and higher), referring to the global
coordinate system in constructing $\overline{a}$ will break rotational
covariance. To remedy this we will depart from ref. \cite{kron} and
introduce a $local$ coordinate system in each cube of the sublattice which
depends on the gauge-invariant behaviour of $U$ on the bounding links of
the cube, as follows.
 The origin of the local coordinates is defined to be
that corner of the cube which has the minimum sum of tr$[U_{\mu \nu}(s) + 
U_{\mu \nu}(s)^{\dag}]$ 
over all plaquettes of the cube which touch the corner. To each direction 
pointing from the local origin to one of the
neighbouring corners of the cube we associate   
tr$[U_{\mu \nu}(s) + U_{\mu \nu}(s)^{\dag}]$ 
of the plaquette of the cube orthogonal to it
and containing the local origin. We arrange these three directions in
ascending order of their associated tr$[U_{\mu \nu} + U_{\mu
\nu}(s)^{\dag}]$ 
and label them as the
local $\hat{1}, \hat{2}, \hat{3}$ respectively. The reader may wish to
ignore rotational invariance on a first reading, and use only the global
coordinates on the whole lattice.

We will define a complete axial gauge inside each cube as follows. First we
define a gauge transformation at each corner of the cube,
\begin{equation}
\Omega[U](z) = U_1(0)^{z_1} U_2(z_1, 0, 0)^{z_2} 
U_3(z_1, z_2, 0)^{z_3}, ~~z_{1,2,3} = 0,1,  
\end{equation}
where we are referring to the $local$ coordinate system.
Then define the gauge transformed lattice field on the bounding links,
\begin{equation}
U  = \overline{U}^{\Omega[U]},
\end{equation}
which satisfies
\begin{equation}
\overline{U}_3 = \overline{U}_2(z_3 = 0) = \overline{U}_1(z_2=z_3=0) = 1.
\end{equation}

Now we interpolate $\overline{U}$ (not $U$!) to each of the bounding 
plaquettes of the cube using the 
2-dimensional interpolation, thereby defining a gauge field 
$\overline{a} \equiv a^{(2)}_{loc-3}[\overline{U}]$ on each bounding plaquette.
An important point here is that the $a^{(2)}$ functional was defined
in the previous subsection by referring to the $global$ coordinate system,
but here we take it to be defined with reference to the $local$ cube
coordinate frame, hence the subscript $loc-3$. 
We smoothly extend $\overline{a}$ to the interior of the cube by defining
$\overline{a}_{\mu}(x)$ to be the solution to the 2-dimensional Laplace
equation on the cross-section of the cube with fixed $x_{\mu}$.

As in 2-dimensions, 
the resulting $\overline{a}$ is not generally transversely 
continuous across boundaries of neighbouring cubes. The cure is to apply a
gauge transformation $\omega$ which ensures that, on the faces of the cube,
the final 3-dimensional
interpolation agrees with the 2-dimensional interpolation of lattice planes
performed in the last subsection, $a^{(2)}[U]$ (where the absence of the
$loc$ 
subscript means that this interpolation was done with reference to the
global coordinate frame). This demand determines $\omega$ on the boundary
of the cube. To see this note that on any cube face, $\overline{a} = 
a^{(2)}_{loc-3}[\overline{U}]$ is gauge equivalent to $a^{(2)}_{loc-3}[U]$
by 2-dimensional gauge covariance. Now $a^{(2)}_{loc-3}[U]$ and 
$a^{(2)}[U]$ differ, if at all,
 only because of the choice of coordinate frame
referred to, different choices being related by an (improper)
two-dimensional rotation. By 2-dimensional rotational invariance, 
$a^{(2)}_{loc-3}[U]$ and $a^{(2)}[U]$ are gauge equivalent. Thus,
$a^{(2)}_{loc-3}[\overline{U}]$ and  $a^{(2)}[U]$ are also gauge
equivalent. The associated gauge transformation, $\omega$, can be readily
determined as the composition of the various transformations associated
with the above lattice gauge transformation and coordinate frame rotation,
as described in the previous subsection.

Now that we have seen that $\omega$ is defined on the cube boundary we note
that it can be extended to the whole cube because the second homotopy class
of $SU(2)$ is trivial (and the boundary of the cube is topologically the same 
as the 2-sphere). We will use the explicit construction that $\omega =
e^{\phi}$
 where $\phi$ is the solution of the 3-dimensional Laplace equation with
boundary condition $\phi = \log \omega$ on the cube boundary.
 Again this extension will $not$ be
continuous if $\omega = -1$ anywhere on the cube boundary because the log
function is discontinuous there, but this situation will only arise for a
measure-zero set of lattice gauge fields which we will neglect.

The final step of the 3-dimensional interpolation is to define
\begin{equation}
a^{(3)}[U] = \overline{a}^{\omega}[\overline{U}].
\end{equation}
It only depends on the link variables, $U$, on the links bounding the cube,
so locality, (iv), is satisfied.

Clearly, $\overline{a}[\overline{U}]$ is a rotationally covariant
functional of $U$, because a rotation of $U$ will induce a rotation of the
$local$ coordinate frame, the frame exclusively referred to in defining 
$\overline{a}$ in 3-dimensions. Since $a^{(3)}[U]$ is gauge equivalent to 
$\overline{a}[\overline{U}]$, it satisfies (ii).

Under a $general$ lattice gauge transformation, $\Omega$, the local
coordinate frame is invariant since its choice was based on the gauge
invariant behavior of $U$. One can easily check that on the links of the
cube, 
\begin{equation}
\overline{U^{\Omega}} = \Omega^{-1}(0) \overline{U} \Omega(0),
\end{equation}
where `$0$' is the origin of local coordinates. As a result,
\begin{equation}
\overline{a}[\overline{U^{\Omega}}] = \Omega^{-1}(0)
\overline{a}[U] \Omega(0), 
\end{equation}
which is a (constant) gauge transformation. It follows that
$a^{(3)}[U^{\Omega}]$ is gauge equivalent to $a^{(3)}[U]$, verifying (iii).

\subsection{4-dimensional interpolation} 

As in three dimensions we define a local coordinate system in each
hypercube of the lattice.
 The origin of the local coordinates is defined to be
that corner of the hypercube which has the minimum sum of tr$[U_{\mu \nu}(s)
+ U_{\mu \nu}(s)^{\dag}]$ 
over all plaquettes of the hypercube which touch the corner. 
 To each direction 
pointing from the local origin to one of the
neighbouring corners of the hypercube we associate the sum of 
tr$[U_{\mu \nu}(s) + U_{\mu \nu}(s)^{\dag}]$
 over all plaquettes of the hypercube orthogonal to it
and containing the local origin. We arrange these four directions in
ascending order of their associated plaquette sums and label them as the
local $\hat{1}, \hat{2}, \hat{3}, \hat{4}$ respectively. 

We again define a lattice gauge transformation on the hypercube to put $U$ 
into a complete axial gauge \cite{luscher},
\begin{equation}
\Omega[U](z) = U_1(0)^{z_1} U_2(z_1, 0, 0, 0)^{z_2} 
U_3(z_1, z_2, 0, 0)^{z_3}  
U_4(z_1, z_2, z_3, 0)^{z_4}, 
\end{equation}
where $z_{\mu} = 0,1$, and we are employing the local coordinate system.
Then define
\begin{equation}
U  = \overline{U}^{\Omega},
\end{equation}
which satisfies
\begin{equation}
\overline{U}_4 = \overline{U}_3(z_4 = 0) = \overline{U}_2(z_3=z_4=0) =
\overline{U}_1(z_2=z_3=z_4=0) = 1.
\end{equation}

Next, interpolate $\overline{U}$ to each of the bounding cubes of the
hypercube using the 
3-dimensional interpolation, thereby defining a gauge field 
$\overline{a} \equiv a^{(3)}_{loc-4}[\overline{U}]$ on each bounding cube.
It is important to note that in the previous subsection the
$a^{(3)}[U]$ functional was defined
with some reference to a $global$ coordinate system, when a gauge 
transformation was performed to ensure agreement with the global
2-dimensional interpolation. By $a^{(3)}_{loc-4}[U]$ we will denote the
3-dimensional interpolation performed when the global coordinate frame is
replaced by the 4-dimensional $local$
coordinates.
We smoothly extend $\overline{a}$ to the interior of the hypercube by defining
$\overline{a}_{\mu}(x)$ to be the solution to the 3-dimensional Laplace
equation on the cross-section of the hypercube with fixed $x_{\mu}$.

Again, the resulting $\overline{a}$ is not generally transversely 
continuous across boundaries of neighbouring hypercubes. We will $try$ to
correct this by applying a 
gauge transformation $\omega$ which ensures agreement with the
interpolation performed in the previous subsection on the 3-dimensional
sublattices of the whole lattice, $a^{(3)}[U]$. 
And again, this demand determines $\omega$ on the boundary of the
hypercube: (a) $\overline{a} = a^{(3)}_{loc-4}[\overline{U}]$ is gauge
 equivalent
to $a^{(3)}_{loc-4}[U]$ by 3-dimensional gauge covariance, and (b)
$a^{(3)}_{loc-4}[U]$ refers to the local hypercube coordinate system while 
 $a^{(3)}[U]$ refers to the global coordinates. The two frames are related by 
a rotation,  so by 3-dimensional rotational covariance, $a^{(3)}_{loc-4}[U]$
and $a^{(3)}[U]$ are gauge equivalent. Therefore $\overline{a}$ and 
$a^{(3)}[U]$ are gauge equivalent on the bounding cubes, and the associated
$\omega$ can be determined as the composition of the various gauge
transformations discussed in the previous subsection, in this case
 associated with the lattice gauge transformation
$\Omega[U]$ and the rotation of the local hypercube frame to the global
frame. 

{\it What makes four dimensions special} is that $\omega$, thusfar defined
only on the boundary of the hypercube, {\it cannot in general} be
continuously extended to the interior of the hypercube. This is because the
third homotopy group of $SU(2)$ is not trivial, but is isomorphic to the
additive group of integers (and the hypercube boundary is topologically the
three-sphere). (Indeed it is this fact that is responsible for the existence
of instantons in four-dimensions.) Let us denote the winding number of
$\omega$: hypercube boundary $\rightarrow SU(2)$,  by $N(s)$,  where $s$ is
the global coordinate of the corner of the hypercube with 
 minimal $s_1 + s_2 + s_3 + s_4$. It can be computed in global coordinates 
using \cite{luscher}
\begin{equation}
N(s) =  - \frac{1}{24 \pi^2} \epsilon_{\mu \nu \alpha \beta} 
\int_{\partial {\rm HC}} dS_{\mu} {\rm tr} 
(\omega^{-1} \partial_{\nu} \omega)(\omega^{-1} \partial_{\alpha} \omega)(
\omega^{-1} \partial_{\beta} \omega),
\label{winding}
\end{equation}
where the integral is over the boundary of the hypercube and $d S_{\mu}$ is
the boundary volume element with an
outward-pointing normal direction associated to it. 
 For a non-zero measure
of lattice gauge fields, there will be hypercubes for which $N(s) \neq 0$,
so that $\omega$ cannot be continuously extended into the hypercube
interior. 

We will proceed by using the following trick. We first choose an 
integer-valued solution to the equation 
\begin{equation}
\sum_{\mu} N_{\mu}(s + \hat{\mu}) - N_{\mu}(s) = N(s),
\label{diver}
\end{equation}
expressed in global coordinates, and associate $N_{\mu}(s)$ to the cube
orthogonal to $\hat{\mu}$, with $s$ being the global coordinates of the 
cube corner with minimal $s_1
+ s_2 + s_3 + s_4$. Before discussing the global existence and choice of such a
solution, let us explain what to do with it. In each cube of the lattice we
define a continuum gauge transformation 
\begin{equation}
\tilde{\omega} \equiv  \omega \sigma^{N_{\mu}(s)},
\label{tilda}
\end{equation}
where in $global$ coordinates,
\begin{equation}
\sigma(s + t_1 \hat{\alpha} + t_2 \hat{\beta} +  t_3 \hat{\gamma}) 
\equiv \frac{(\sum_a \tau_a {\rm cot}(\pi t_a) + i \epsilon_{\mu \alpha
\beta \gamma})^2}{
\sum_a {\rm cot}^2(\pi t_a) + 1}, 
\end{equation} 
and $\alpha < \beta < \gamma$ are the global frame
directions orthogonal to $\hat{\mu}$, and $0 \leq t_a \leq 1$.
 Note that $\sigma = 1$ on the
boundaries of each cube and wraps around $SU(2)$ once \footnote{This map
was obtained by composing a map from the unit cube onto infinite 
3-dimensional space followed by a map from this space
 onto $SU(2)$. The first map is obtained by
1-dimensional sterographic projection of the unit interval (thought of as a
circle) onto the real line. The second map is obtained by the inverse of
3-dimensional stereographic projection (recall that $SU(2)$ is a 3-sphere).}.
The $\sigma^{N_{\mu}(s)}$ define a
continuous map on the boundary of each hypercube, which we will denote by
$\tilde{\sigma}$. 
It may be readily
verified that it has a winding number of 
$\sum_{\mu} N_{\mu}(s) - 
N_{\mu}(s + \hat{\mu}) = - N(s)$. Therefore $\tilde{\omega}$
is also a continuous map on a hypercube boundary, which clearly has
trivial winding, so it can be extended to the whole hypercube.
We will choose this extension to be the one which minimizes the action
\begin{equation}
S = {\rm tr} \int d^4x ~~\partial_{\mu} \tilde{\omega}^{-1} \partial_{\mu}
 \tilde{\omega}
\label{min}
\end{equation}
in each hypercube.

Our final interpolation will be defined as
\begin{equation}
a[U] = \overline{a}^{\tilde{\omega}}[\overline{U}].
\end{equation}
Note that on the cubes bounding the hypercube,
\begin{equation}
a[U] = (a^{(3)}[U])^{\tilde{\sigma}},
\end{equation}
so $a[U]$ is transversely continuous across hypercube boundaries.

Now let us return to the question of the global existence and uniqueness of 
$N_{\mu}(s)$ on a  toroidal lattice. We are representing this lattice by 
global coordinates $s: 0 \leq s_{\mu} \leq L$ where we will eventually wish to
identify $s_{\mu} = L$ with $s_{\mu} = 0$. We will demand solutions to
eq.(23) which satisfy the boundary conditions,
\begin{eqnarray}
N_{j}(s_j = L) = N_j(s_j = 0), ~~ j= 1,2,3, \nonumber \\
N_{4}(s_4 = L) = N_{4}(s_4 = 0) + \delta_{0 s_1} \delta_{0 s_2} \delta_{0
s_3} N[U],
\end{eqnarray}
where 
\begin{equation}
N[U] \equiv \sum_{{\rm hypercubes}} N(s).
\end{equation}
Thus $N_{4}$ is $not$ always periodic in the $\hat{4}$-direction.

$N_{\mu}$ solutions are not unique because of the invariance of
eq.(23) under the `gauge transformation' 
\begin{equation}
N_{\mu} \rightarrow N_{\mu} + \epsilon_{\mu \nu \alpha \beta} \{N_{\alpha
\beta}(s + \hat{\nu}) - N_{\alpha \beta}(s)\},
\end{equation}
where $N_{\alpha \beta}(s)$ is any integer-valued function associated with
each plaquette. In order to have a definite interpolation algorithm we will
pick a `gauge-fixing' scheme for the $N_{\mu}$ on the 
lattice. For example, choose a path passing through each lattice hypercube
once, starting from the hypercube at $(0,0,0,L-1)$, and not crossing the 
boundaries $x_{\mu} = 0, L$. Demand that
$N_{\mu} = 0$ on all cubes {\it not} intersected by the path,
 with the exception that
$N_4(0,0,0,L) = N[U]$ (see eq.(29)). This effectively provides a complete
 gauge-fixing scheme, compatible with the boundary conditions on
$N_{\mu}$.

With our boundary conditions it is easy to see that unless $N[U] = 0$,
the gauge field interpolation $a[U]$ is not fully periodic, but rather is
periodic modulo a gauge transformation,
\begin{equation}
a_j(x_4 = L) = a_j^{\tilde{\sigma}}(x_4 = 0).
\end{equation}
Now $\tilde{\sigma}$ winds $N[U]$ times around $SU(2)$ on the boundary $x_4 =
L$. This is nothing but a boundary condition for a continuum gauge field representing an $SU(2)$ connection with winding number,
\begin{eqnarray}
 N[U] = - \frac{1}{16 \pi^2} \int d^4 x Tr[ f_{\mu\nu}\; \tilde{f}_{\mu\nu} ],
\end{eqnarray}
where $f_{\mu\nu}$ is the field strength of the continuum field $a_\mu$.
This definition of winding number is very closely related to that given by 
L\"uscher \cite{luscher} \footnote{In fact, the only difference is in the particular interpolation of $\omega$ used.}. In the event that $N[U] = 0$ it is easy to
see that $a[U]$ is $fully$ periodic. Only such gauge fields are required in
our earlier proposal for formulating chiral gauge theories \cite{us}, as
mentioned in the introduction.

As in three dimensions it is clear that $\overline{a}[\overline{U}]$ is a
rotationally and translationally covariant functional of $U$. Since $a[U]$
is gauge equivalent to $\overline{a}[\overline{U}]$ it must also satisfy
(ii).
And also as in three dimensions the fact that we took $U$ to its completely
gauge-fixed form, $\overline{U}$, prior to interpolating, ensures that
gauge covariance, (iii), holds.

While it is clear that $\overline{a}[\overline{U}]$ is determined only by the 
$U$ on the bounding links of the hypercube,  $a[U]$ is $not$
a local functional of $U$ in general because $\tilde{\omega}$ is determined
$locally$ in terms of the fields $U$ $and$ $N_{\mu}(s)$, but  
 $N_{\mu}(s)$   depends $non-locally$ on $U$ through eq. (23). Nevertheless
property (iv) is satisfied because  the trace of 
any Wilson loop determined by $a[U]$ can be broken up into a product of 
segments that are each contained in a single hypercube, of the form
\begin{eqnarray}
P~e^{i \int_{x_0}^{x_1} d x^{\mu}  a_{\mu}[U](x)} &=&
\tilde{\omega}^{-1}(x_0)~ P~e^{i \int_{x_0}^{x_1} d x^{\mu}  
\overline{a}_{\mu}[\overline{U}](x)}~
\tilde{\omega}(x_1) \nonumber \\
&=& \tilde{\sigma}^{-1}(x_0) \omega^{-1}(x_0)~
P~e^{i \int_{x_0}^{x_1} d x^{\mu}  \overline{a}_{\mu}}~
\omega(x_1) \tilde{\sigma}(x_1),
\end{eqnarray}
where the first equality follows from $a[U] =
\overline{a}^{\tilde{\omega}}$ and the second from the fact that 
$\tilde{\omega} = \omega \tilde{\sigma}$ on hypercube boundaries (where
$x_{0,1}$ lie). In the
trace of the product of such segments making up a Wilson loop, the
$\tilde{\sigma}$ dependence cancels out. Because $\overline{a}$ and
$\omega$ in any hypercube are determined only by $U$ on the bounding links
of the hypercube, (iv) follows.

\section{The example of compact $U(1)$ in two dimensions}

One may also wonder how to treat $U(1)$ factors of the gauge group when
they are taken as compact\footnote{The non-compact case is much simpler, see
\cite{us} for an explicit 4-D interpolation.}. 
For this case, our construction breaks down in
two dimensions because we cannot in general extend $\omega$ defined on the
plaquette boundaries to the whole plaquette since $U(1)$ is not simply
connected. In two-dimensions the problem is easily solved by constructing the
analog of the $\tilde{\omega}$ map, needed in four dimensions for the
non-abelian case. In three or four-dimensions we note that the compactness
of $U(1)$ is only relevant in the continuum limit if the $U(1)$ is the
result of spontaneous breaking of a non-abelian symmetry. Therefore one can
handle this case by keeping the original non-abelian gauge group, and
whatever matter fields lead to its spontaneous breakdown. The non-abelian
lattice gauge fields are then to be treated by the methods of the previous
section.

It is instructive to construct the explicit expression for the interpolation 
in compact QED in two dimensions, to illustrate some of the methods in
section 2 in a simple setting,  and to 
 compare the method with previous proposals in the 
literature.  
According to the procedure of subsection 2.2, we get 
\begin{eqnarray}
\overline{a}_{1}(s + t_1 \hat{1} + t_2 \hat{2}) &=&  - i \;t_2 \log(U_{12}(s)) 
\nonumber\\
\overline{a}_{2}(s + t_1 \hat{1} + t_2 \hat{2}) &=& 0,
\end{eqnarray}
while $\omega$ on the links is,
\begin{eqnarray}
\omega(s + t_1 \hat{1}) &=& e^{i t_1 A_{1}(s)}, \nonumber \\
\omega(s + \hat{1} + t_2 \hat{2}) &=& e^{i A_{1}(s) + i t_2 
A_{2}(s + \hat{1})}, 
\nonumber \\
\omega(s + t_2 \hat{2}) &=& e^{i t_2 A_{2}(s)}, \nonumber \\
\omega(s + t_1 \hat{1} + \hat{2}) &=& e^{- t_1 \log U_{12}(s)+ 
i A_{2}(s) + i t_1 A_{1}(s + \hat{2})},
\end{eqnarray}
Using the two dimensional analog of eq. (\ref{winding}),
\begin{eqnarray}
N(s) = \frac{i}{2 \pi} \epsilon_{\mu\nu} \int_{\partial {\rm P}} 
d S_\mu \; \omega^{-1} \partial_\nu \omega 
\end{eqnarray}
we find,
\begin{eqnarray}
N(s) = Int[\;\chi(s), 2\pi]
\end{eqnarray}
where $Int[,2 \pi]$ denotes the nearest integer part modulo $2 \pi$ and
\begin{eqnarray}
\chi(s) \equiv A_{2}(s) + A_{1}(s+\hat{2}) - A_{2}(s+\hat{1}) - A_{1}(s)
\end{eqnarray}

 In general a non-measure zero set of  
lattice configurations will have $N(s) \neq 0$ and in this case $\omega$ cannot 
be extended smoothly to the interior of the plaquette.  According to the 
rules of 
section 2.4, in order to proceed we must solve the 
integer equation (\ref{diver}) in the `gauge' depicted in Fig. 1, where we
take our original spacetime to be a torus. 
All the links that are not crossed by the path have the associated
$N_{\mu}(s)$ set to zero (remember that $\hat{\mu}$ is $orthogonal$ to the
link in our notation). 
The $N_{\mu}(s)$ solution  is then clearly unique (the explicit closed form
expression is not very illuminating, and so is omitted).
 It is easy to check  that the gauge transformation defined on each link by
\begin{eqnarray}
\tilde{\sigma}(s +  t_{\nu} \hat{\nu})
 \equiv e^{i\; 2 \pi
\epsilon_{\mu\nu} N_{\mu}(s) \; t_{\nu}}
\end{eqnarray}
has winding number $-N(s)$ and that $\tilde{\omega}$ defined in (\ref{tilda}) 
has  zero winding. Thus it  
can be extended to the interior of the plaquette. It is easy to  
explicitly find the minimum of equation (\ref{min}) in this case,  
\begin{eqnarray}
\tilde{\omega}(s+ t_1 \hat{1} + t_2 \hat{2}) = e^{i \;\phi(t_1,t_2)}, 
\end{eqnarray}
with
\begin{eqnarray}
\phi(t_1,t_2) = (A_1(s) - 2 \pi N_2(s)) \;t_1 + ( A_2(s) + 2 \pi N_1(s) )
\;t_2 \nonumber\\
+ (A_2(s+\hat{1}) + 2 \pi N_1(s+\hat{1}) - A_2(s) - 2 \pi N_1(s)) \; t_1 t_2  
\end{eqnarray}  
and the final expression for ${\overline{a}}^{\tilde{\omega}}(s+t_1
\hat{1}+t_2\hat{2})$ is, 
\begin{eqnarray}
a_1 &=& (1 -\; t_2)\; ( A_1(s) - 2 \pi N_2(s) ) + t_2
\; ( A_1(s+\hat{2}) - 2 \pi N_2(s+\hat{2}))\nonumber\\
a_2 &=& (1 -\; t_1) \;( A_2(s) + 2 \pi N_1(s) )
+ t_1 \; ( A_2(s+\hat{1}) + 2 \pi N_1(s+\hat{1}) ) 
\label{interqed}
\end{eqnarray}
It is important to point out that even though the interpolated gauge fields are
non-local, due to the $N_{\mu}$ ``fields'', the 
field strength of (\ref{interqed}) is local, and equal to $- i 
\log(U_{12}(s))$. Consequently, all gauge invariant 
quantities (which in $QED_2$ can be constructed from the field strength) 
are local. Also, as expected the interpolation is transversely continuous and
covariant  under lattice gauge transformations, rotations and 
translations.

This interpolation has important differences with previous interpolations 
in the literature. The interpolation
in ref. \cite{kron} for compact $QED_2$ differs in that  
the authors directly interpolate $\omega$ 
instead of $\tilde{\omega}$, thus getting singular gauge fields whenever 
any $N(s)$ is non-zero. Such gauge fields are unsuitable for our
formulation of chiral gauge theories \cite{us}.
 On the other hand, Flume and Wyler in ref. \cite{int1} get 
a smooth interpolation, but at the expense of breaking $compact$ gauge 
covariance. This 
is also the case in the last reference of \cite{int2}.  

\section{Concluding Remarks}

In ref. \cite{us} we actually needed an interpolation of $U$ to a $finer$
lattice, rather than the continuum. This is easily accomplished by taking
$\overline{a}$ to live on the links of the finer lattice, and
$\tilde{\omega}$ to live on the vertices of the fine lattice and replacing
the various continuum Laplace equations by lattice equations on the fine
lattice. The interpolated link variables $u_{\mu}$ are then given by
\begin{equation}
u_{\mu}(x) = \tilde{\omega}^{-1}(x) e^{i f \overline{a}_{\mu}(x)} \tilde{
\omega}(x + f
\hat{\mu}),
\end{equation}
where $f$ is the lattice spacing for the fine lattice. The integral formula
for $N(s)$ in terms of $\omega$ can be replaced by the lattice equivalent,
rounded to the nearest integer. The resulting interpolation will agree with
the continuum interpolation in the limit $f \rightarrow 0$. 

One of the standard tests of success for any scheme which claims to
preserve chiral symmetries in the continuum limit is to look at what
happens at exactly $g = 0$, where $g$ is the gauge coupling. The
exponential of the Wilson action for gauge fields becomes a $\delta$-function
which only permits lattice gauge fields which are pure gauge.  The test is
then to see if the fermions coupled to the gauge field become free chiral
fermions at $g = 0$ in the continuum limit (see for example \cite{golt}).
It is interesting to see how this works in our formulation of lattice
chiral gauge theory \cite{us}. In our interpolation procedure, it is
straightforward to verify that in the special case $U = 1$ everywhere, the
interpolation is just $a_{\mu} = 0$ everywhere. So by property (iii) of gauge
covariance, when $U$ is pure gauge, the interpolation is a transversely
continuous gauge field which is pure gauge.
 In ref. \cite{us} we showed that in our
continuum limit the fermions are  gauge-invariantly coupled to transversely
continuous gauge field interpolations. 
In particular therefore, when the interpolation is pure gauge
the fermions are free.

\begin{figure} 
\begin{center}
\mbox{
\epsfig{file=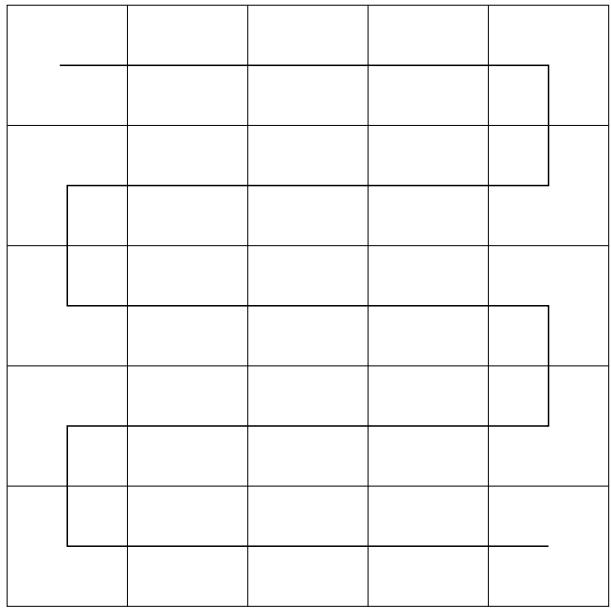,height=8cm}}
\end{center}
 \caption[]{`Gauge choice' used in solving (\ref{diver})
for compact $QED_2$. The $N_{\mu}$ corresponding to links
 that are not intersected by the path are zero.}
 \label{fig:lattice}
\end{figure}

\section*{Acknowledgments}

We wish to thank R. Narayanan for helpful criticism of
our earlier interpolation scheme, and J. L. Alonso, Ph. Boucaud and A. van
der Sijs for discussions. 
This research was supported by NSF-PHYS-92-18167. P.H. is supported
by the Harvard Society of Fellows and also partially by the  
Milton Fund of Harvard University.

\end{document}